\documentclass[twocolumn,preprintnumbers,amsmath,amssymb]{revtex4-1}
\usepackage{graphicx}
\usepackage{dcolumn}
\usepackage{bm}
\usepackage{latexsym}
\usepackage{xcolor}
\usepackage{changes}
\newcommand{\RN}[1]{%
  \textup{\uppercase\expandafter{\romannumeral#1}}%
}
\begin{document}
\title{A biologically motivated three-species exclusion model: effects of leaky scanning and overlapping genes on initiation of protein synthesis}
\author{Bhavya Mishra}
\affiliation{Department of Physics, Indian Institute of Technology Kanpur, 208016}
\author{Debashish Chowdhury}
\email[E-mail:]{debch@iitk.ac.in}
\affiliation{Department of Physics, Indian Institute of Technology Kanpur, 208016}

\begin{abstract}
Totally asymmetric simple exclusion process (TASEP) was originally introduced as a model for the traffic-like collective movement of ribosomes on a messenger RNA (mRNA) that serves as the track for the motor-like forward stepping of individual ribosomes. In each step, a ribosome elongates a protein by a single unit using the track also as a template for protein synthesis.  But, pre-fabricated, functionally competent, ribosomes are not available to begin synthesis of protein; a subunit directionally scans the mRNA in search of the pre-designated site where it is supposed to bind with the other subunit and begin the synthesis of the corresponding protein. However, because of `leaky' scanning, a fraction of the scanning subunits miss the target site and continue their search beyond the first target. Sometimes such scanners successfully identify the site that marks the site for initiation of the synthesis of a different protein. In this paper, we develop an exclusion model, with three interconvertible species of hard rods, to capture some of the key features of these biological phenomena and study the effects of the interference of the flow of the different species of rods on the same lattice. More specifically, we identify the meantime for the initiation of protein synthesis as appropriate mean {\it first-passage} time that we calculate analytically using the formalism of backward master equations. In spite of the approximations made, our analytical predictions are in reasonably good agreement with the numerical data that we obtain by performing Monte Carlo simulations. We also compare our results with a few experimental facts reported in the literature and propose new experiments for testing some of our new quantitative predictions. 
\end{abstract}

\maketitle
\section{Introduction}
The asymmetric simple exclusion process (TASEP) \cite{evans05,blythe07,schutz01,derrida98,mallick15} is the simplest model of interacting self-propelled particles. In its simplest version, under open boundary conditions, particles enter from one end of a one-dimensional lattice and exit from the other end after traversing the distance between the two ends by a sequence of hops from one lattice site to the next; however, a successful hop is allowed if, and only if, the target site is not already occupied by another particle. Fundamental properties of TASEP, like its phase diagram in the non-equilibrium steady state (NESS), are determined by the three key parameters, namely, the rates of entry ($\alpha$), exit ($\beta$) and forward hopping in bulk ($q$) of the lattice \cite{krug91,derrida98,schutz01,blythe07}. Many extensions of this model have found applications in modeling wide varieties of collective traffic-like phenomena that occur at many levels- from molecules to vehicles \cite{chowdhury00,schad10,chowdhury05,chou11,chowdhury13a,rolland15,
macdonald68,macdonald69,tripathi08,klumpp08,klumpp11,sahoo11,
ohta11,wang14,schutz03,kunwar06,john04,lin11,lakatos03,shaw03,
shaw04a,shaw04b,chou03,chou04,zia11,chou99,levine04,liu10,
ciandrini10,basu07,garai09,greulich12,kuan16,chowdhury08,
oriola15,sugden07,evans11,chai09,ebbinghaus09,ebbinghaus10,
muhuri10,neri11,neri13a,neri13b,curatolo16,klein16,parmeggiani04,graf17}. 

Interestingly, TASEP was introduced initially fifty years ago as a model for the synthesis of biopolymers, called protein, by macromolecular machines called ribosome \cite{macdonald68,macdonald69}. In that formulation, the messenger RNA (mRNA), that serves as the template for protein synthesis and also as the track for the motor-like movement of the ribosomes, is represented by a lattice each site of which, labelled by the integer index $j$ ($1 \leq j \leq N$), is called a codon. Since a ribosome is, typically, ten times longer than a codon, the authors of \cite{macdonald68} denoted each ribosome by a hard rod of length ${\ell}>1$, instead of representing these by particles (${\ell}=1$), where length of both the lattice ($N$) and the rods (${\ell}$) are measured in the unit of lattice spacing. Since many ribosomes can simultaneously move co-directionally on the same mRNA track, each synthesizing a distinct copy of the same protein, their traffic-like collective movement can be modeled as a TASEP of hard rods.  

In the minimal model, entry of the rods onto the lattice through the site $j=1$, which is identified as the start codon,  captures all the chemo-mechanical processes involved in the {\it initiation} of protein synthesis regarding a single effective rate $\alpha$.  Consequently, many exotic phenomena associated with protein synthesis are beyond the scope of the minimal TASEP-based model described above. One such biophysical phenomenon, that arises from the complexities of the {\it initiation} of protein synthesis, motivated our study the results of which are reported here. 

First, unlike macroscopic machines, pre-fabricated, functionally competent, ribosomes are not available off the shelf to begin synthesis of protein. One of the major components of a ribosome called small subunit (SSU), first binds an arbitrary site upstream from the start codon on the mRNA. Then it searches for the specific start codon on that mRNA by scanning along the latter's contour \cite{topis11,kozak99,jackson10,hinnebusch14,hinnebusch16,hinnebusch17,hinnebusch11,alasek12} in a directed manner \cite{vassilenko11a, chu12, vassilenko11b, andreev18} although the source of the energy required for directed motion is still debated \cite{kozak78,kozak80,kozak89}. Once the SSU reaches the start codon and binds it specifically, a LSU binds with it and with the mRNA at the start codon, along with some other accessory molecules, the assembly of the ribosome is complete, and protein synthesis can begin. 

Second, further intricacies arise from the fact that more than one start codon may exist on an mRNA indicating the possibility of synthesis of more than one distinct species of protein molecules using different segments of the same mRNA. Loosely, using the terminology of biology, we call the segment of mRNA in between a start codon and the corresponding stop codon as a `gene.' If two genes (from now onwards identified as `Gene 1' and `Gene 2'),  encoded on the same mRNA overlap, then the rates of synthesis of the corresponding proteins (from now onwards referred to as `P1' and `P2', respectively) would suffer because of the interference of the traffic of the two species of ribosomes engaged in their synthesis. 

Motivated by the phenomena outlined above, we develop here a 3-species exclusion model. In the language of TASEP, we study here an exclusion model with {\it three} species of rods where each member of the species rod0 represents a scanning SSU. At the first of the two start sites, a fraction of the rod0 gets converted into the species rod1 mimicking the binding with an LSU thereby completing the assembly of a ribosome competent to begin synthesis of a protein P1. A rod0 may miss the first start site and unknowingly continue its search for a start site.  Those rod0 which miss the first start site because of `leaky' scanning \cite{michel14,white04,kozak02}, find themselves in a co-directionally moving traffic of rod1 and thereby interfere with the synthesis of P1 till they reach the second start site. A fraction of these surviving rod0 correctly identify the second start site and get converted into rod2 thereby becoming competent to begin synthesizing a protein P2. The remaining fraction of rod0 that fail to identify also the second start site continues scanning till it finally detaches from the lattice after reaching a particular site. 

Depending on the positions of the two pairs of start and stop sites, the flow of all the three species may interfere over a segment of the lattice whereas only two of the three species may interfere elsewhere on the same lattice. In this paper, we focus on the consequences of these interferences of the exclusion processes on the rates of initiation of the synthesis of the two proteins P1 and P2. We formulate the process of reaching the start sites through the scanning-based search process as {\it first-passage} problems (see ref. \cite{feller08, gardiner03,redner01,redner14,bressloff14,chou14} for {\it first passage} problem). 

Formulation of any first-passage problem requires the unambiguous specification of the initial and final states. For our purpose in this paper, we first allow the system to evolve according to the kinetic rules of the model (stated in the next section). Long after the system attains a non-equilibrium steady state, characterized by a constant average flow of the three species of rods, we switch on the clock, setting it to $t=0$, and begin monitoring the free rods in the surrounding pool (all of which are of the same species rod0) with the passage time. Because of the possibility of inter-species conversion, the target sites need not be hit by the very first rod that enters through the first site on the lattice. But, as soon as a rod0 hits the first or the second target site the clock is stopped and the corresponding time displayed on the clock is recorded as the corresponding first-passage time (FPT). Then the whole process is repeated for recording the next  FPT. Thus the first-passage problem studied here refers not to that of any particular rod but a pool of rods none of which is attached to the lattice at $t=0$; the duration that elapses before the first successful hitting of the target by any of the members of that pool is identified as the FPT.  

In our Monte Carlo (MC) simulations, this algorithm is implemented verbatim and the data collected over a large number of {\it in-silico} experiments are averaged to get the two mean-first passage times (MFPTs). We calculate the same MFPTs also analytically utilizing the formalisms of backward master equations under mean-field approximation. In spite of the approximations made in the mean-field theory, we find reasonably good agreement between the theoretical predictions and MC data. Our results also account for some experimental data, and we make new, experimentally testable, theoretical predictions. 

\section{Model}

\begin{figure}[ht]
\begin{center}
\includegraphics[width=0.95\columnwidth]{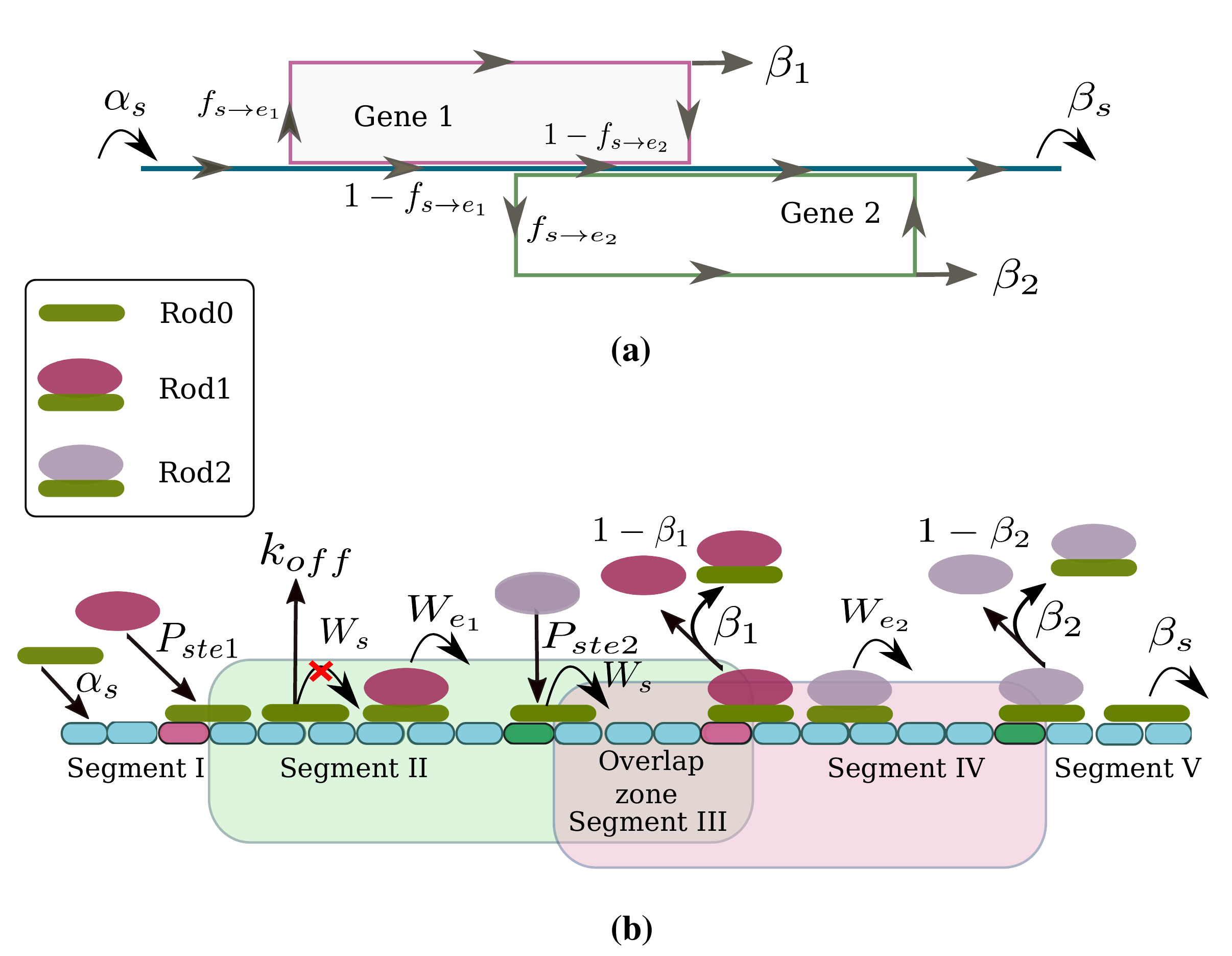}
\end{center}
\caption{(Color online) A schematic description of 3-species exclusion process. In (a) we show the fraction of scanning SSU and the two groups of ribosomes elongating proteins encoded by Genes 1 and 2. In (b) the mRNA is depicted as a long one-dimensional lattice of length $L$, with five distinct segments labeled by I-V. On this lattice, three species of rods, named rod0, rod1 and rod2, hop from left to right, respecting the constraint of mutual exclusion, with the respective rates next to the arrows that indicate the corresponding process (see the text for detailed description of the processes and the five different segments of the lattice). 
}
\label{fig-model_1}
\end{figure}

Our model is shown schematically in Fig.\ref{fig-model_1}. In this model, mRNA is treated as a long one-dimensional lattice of length $ N $ and ribosomes as the hard rod of length $ \ell $, in units of lattice site spacing. Here, each single lattice site represents a codon. The segment of the lattice between a start site and stop site is identified as a `gene.' For the sake of simplicity, we include only two overlapping genes in our model. Accordingly, there are three distinct species of rods in our model; one of these species represent scanning SSUs while the other two species of rods represent two groups of ribosomes engaged in the synthesis of proteins as directed by the two genes $ 1 $ and $ 2 $. For brevity, we'll use the terms rod0, rod1, and rod2 to denote these three species of rods.  
The model can be easily extended to situations where $G$ genes overlap on a single mRNA; the number of species of rods required would be at least $G+1$. Again for simplicity, we assume all the rods to have the same length although, in reality, the length of a SSU can be slightly different from those of the fully assembled ribosomes. 

In our notation lattice sites are labelled from left to right with the integer index $ i=1,2,\dots,N+\ell-1 $. The position of a ribosome along the lattice is denoted by the site covered by the leftmost end of the corresponding rod. At any instant of time $ t $, if a rod is located at site $ i $, i.e. its leftmost edge is at site $ i $, then site $ i $ is said to be `occupied' and all remaining sites from $ i+1,i+2,\dots,i+\ell-1 $ are said to be `covered' by the same rod simultaneously. Mutual exclusion constraint among the ribosomes is imposed by the condition that no two rods can occupy/cover the same lattice site simultaneously.

Although ${\ell} > 1$ (typically ${\ell}=10$), the rod is allowed to hop forward by only one lattice site at a time, thereby mimicking the codon-to-codon stepping of a ribosome; in each step, it elongates the nascent protein by a single subunit. 

On the basis of location of start and stop codons of gene $ 1 $ and gene $ 2 $, we divide the lattice into five segments 
\begin{eqnarray}
\RN{1}&:& 1 \leq i \leq n_{i1} \nonumber \\ 
\RN{2}&:& n_{i1}+1 \leq i \leq n_{i2}  \nonumber \\ 
\RN{3}&:& n_{i2}+1 \leq i \leq n_{s1}  \nonumber \\ 
\RN{4}&:& n_{s1}+1 \leq i \leq n_{s2}  \nonumber \\
\RN{5}&:& n_{s2}+1 \leq i \leq N,   
\end{eqnarray}
where the lattice sites $n_{i1}$ and $n_{i2}$ correspond to the start (initiation) codons of the genes 1 and 2, respectively, from where the synthesis of the respective proteins initiate, while  $n_{s1}$ and $n_{s2}$ mark the corresponding stop sites.

The complete set of rules for the kinetics of hopping of the three species of rods in our model are given as follows:

(i) A new rod0 can attach at site $ i=1 $  with rate $ \alpha_s $, provided all $ \ell $ consecutive sites from $ i=1 $ to $ i=1+\ell-1 $ are empty.

(ii) If there is a rod0 occupying site $ i=N $, it can detach from that lattice with rate $ \beta_s $.

(iii) If there is a rod0 occupying site $ i=n_{i1} $ i.e. the start codon of gene $ 1 $, then it can 

(a) either hop forward with the composite rate $P_{ste1} \times W_{e1}$ (i.e., transform into rod1, with the probability $P_{ste1}$ and hop forwared by one step to the site $i+1$, as rod1, with rate $W_{e1} $), provided the target site $i+1$ is not covered by any other rod),

(b) or hop forward with the composite rate $(1-P_{ste1}) \times W_{s}$  (i.e., remain a rod0, with the probability $1-P_{ste1}$ and hop forward by one step to site $ i+1 $ as rod0 only, with rate $ W_s $), provided the target site $i+1$ is not already covered by any other ribosome).

(iv) If there is a rod0 occupying site $ i=n_{i2} $ i.e. the start codon for gene $ 2 $, then it can 

(a) either hop forward with the composite rate $P_{ste2} \times W_{e2}$ (i.e., transform into rod2, with the probability $P_{ste2}$ and hop forwared by one step to the site $i+1$, as rod2, with rate $W_{e2} $), provided the target site $i+1$ is not covered by any other rod),

(b) or hop forward with the composite rate $(1-P_{ste2}) \times W_{s}$  (i.e., remain a rod0, with the probability $1-P_{ste2}$ and hop forward by one step to site $ i+1 $ as rod0 only, with rate $ W_s $), provided the target site $i+1$ is not already covered by any other ribosome).

(v) If a rod1 is occupying the site $ i=n_{s1} $ i.e. the stop codon of gene $ 1 $, then it could either detach from the lattice with rate $ \beta_1 $ or transform into rod0, and further hope forward by one step with rate $ W_{s} $, provided the target site is not covered by any other rod.

(vi) If rod2 is occupying the site $ i=n_{s2} $ i.e. the stop codon of gene $ 2 $, then it could either detach from the lattice with rate $ \beta_2 $ or transform into rod0, and further hop forward by one step with rate $ W_{s} $, provided the target site is not covered by any other rod.

(vii) At all other sites, if the site is occupied either by a rod0, rod1 or rod2, the rod can hop forward with their respective rates $ W_s, W_{e1} $ and $W_{e2}$, provided the target site is not occupied by any other rod. 

(viii) A rod0 can detach from the lattice, with the rate $K_{off}$ if it occupied a site $j$ in segment II, provided there is a rod1 immediately in front of it, i.e., occupying the site $j+{\ell}$. 

Throughout the lattice rods follow the exclusion constraint, i.e., no two rods can occupy the same lattice site simultaneously.
Drawing analogy with the phenomenon of translation by ribosomes, which has motivated this study, we refer to the segments I and II as the untranslated regions 1 and 2 (UTR$_{1}$ and UTR$_{2}$), respectively, for the genes that are captured by the start sites $i=n_{i1}$ and $i=n_{i2}$, respectively. 


\section{Results and discussion}

In Monte-Carlo simulations, we adopt the random sequential updating scheme to measure the effect of crowding of rods on the MFPTs. In order to convert an arbitrary rate constants $k$ into the corresponding dimensionless probability $p_k$, we apply the formula $ p_{k}=k~dt $, where $ dt $ is a sufficiently small time interval. The  typical
numerical  value  of $ dt $ used in our simulations is $ dt=0.01s $. Unless stated explicitly otherwise, the numerical values of the relevant parameters used in the simulations are  $ \beta_s=0.8, P_{ste2}=0.7,W_s=1.0,W_{e1}=1.0,W_{e2}=1.0,\beta_1=0.8,\beta_2=0.8,~\text{and}~\ell=10 $.

\subsection{Mathematical Formulation}

We use the symbol $\rho_{\lambda,\mu}$ is the average number density of rods species $\lambda$ 
($\lambda=0,1,2)$ in the segment $\mu$ ($\mu=I,II,III,IV,V$) of the lattice. Therefore 
$\rho_{\mu} = \sum_{\lambda=0}^{2} \rho_{\lambda,\mu}$ is the total number density in the segment $\mu$ irrespective of the species of the rod. 
Suppose $ P(\underbrace{0,0, \dots, 0}_{\ell}) $ denotes the probability that all consecutive $ \ell $ sites from $ i=1 $ to $ i=1+\ell-1 $ are empty, simultaneously. In steady state the expression of $P(\underbrace{0,0, \dots, 0}_{\ell}) $ is given by
\begin{equation}
P(\underbrace{0,0, \dots, 0}_{\ell})=(1-\rho_{0,I})^\ell.
\label{cond_2}
\end{equation} 
where $\rho_{0,I}$ is  the number density of rod0 in segment I.\\

We define $ q_{\mu,\nu}(\underline{i}|i+\ell,t) $  as the conditonal probability at time $t$ that the site $ i+\ell $ in segment $ \nu$ is not occupied by any {\bf rod}, given that there is a {\bf rod} at site $ i $ in segment $ \mu $. 
In the steady state the expression of $ q_{\mu,\nu}(\underline{i}|i+\ell,t) $ is given by \cite{basu07,garai09},
\begin{equation}
q_{\mu,\mu}(\underline{i}|i+\ell,t)=\dfrac{1-\rho_{\mu}\ell}{1+\rho_{\mu}-\rho_{\mu}\ell},
\label{cond_1}
\end{equation}
where $\rho_{\mu}$ is the total number density in the segment $\mu$ (i.e., number density irrespective of the species of rods).

Next, we define mean first passage times $ \langle t_{1} \rangle $, and $\langle t_{2} \rangle$ as the average times taken by a rod0 to reach the start sites $j=n_{i1}$ and $j=n_{i2}$, respectively, for the first time, if it was initially in state $ j=0 $ (i.e. in the free rod0 pool at time $ t=0 $). We apply the method of backward Master equations \cite{kolomeisky11,veksler13,kolomeisky12,shvets16} to get the analytical expressions of 
$\langle t_{1} \rangle$ and $\langle t_{2} \rangle$. 

Define, $ F_{n}(m,t) $ as a first passage probability to find a ribosome occupying site $ m $ at time $ t $ if it was at site $ n $ at time $ t=0 $. The time evolution of $ F_{n}(m,t) $ lead us to the set of backward master equations listed below. \\
At site $ n=0 $,
\begin{eqnarray}
\dfrac{dF_{0}((m=n_{i1},n_{i2}),t)}{dt}=\alpha_{s}P(\underbrace{0,0, \dots, 0}_{\ell})F_{1}(m,t) \nonumber \\ 
-\alpha_{s}P(\underbrace{0,0, \dots, 0}_{\ell})F_{0}(m,t).\nonumber \\
\label{m_eq_0}
\end{eqnarray}
For $ 1 \leq n<n_{i1}-\ell $,
\begin{eqnarray}
\dfrac{dF_{n}((m=n_{i1},n_{i2}),t)}{dt}=W_sq_{I,I}(\underline{n}|n+\ell)F_{n+1}(m,t) \nonumber \\ 
-W_sq_{I,I}(\underline{n}|n+\ell)F_{n}(m,t).\nonumber \\
\label{m_eq_11}
\end{eqnarray}
For $ n_{i1}-\ell \leq n<n_{i1} $,
\begin{eqnarray}
\dfrac{dF_{n}((m=n_{i1},n_{i2}),t)}{dt}=W_sq_{I,II}(\underline{n}|n+\ell)F_{n+1}(m,t)\nonumber \\  -W_sq_{I,II}(\underline{n}|n+\ell)F_{n}(m,t),\nonumber \\
\label{m_eq_12}
\end{eqnarray}
At site $ n=n_{i1} $,
\begin{eqnarray}
\dfrac{dF_{n}(m=n_{i2},t)}{dt}=W_sq_{I,II}(\underline{n}|n+\ell)F_{n+1}(m,t)\nonumber \\ -W_sq_{I,II}(\underline{n}|n+\ell)F_{n}(m,t).\nonumber \\
\label{m_eq_2}
\end{eqnarray}
For $  n_{i1}+1 \leq n<n_{i2}-\ell $,
\begin{eqnarray}
\dfrac{dF_{n}(m=n_{i2},t)}{dt}=W_s(1-P_{ste1})q_{II,II}(\underline{n}|n+\ell)F_{n+1}(m,t)\nonumber \\ -W_s(1-P_{ste1})q_{II,II}(\underline{n}|n+\ell)F_{n}(m,t).\nonumber \\
\label{m_eq_3}
\end{eqnarray}
For $  n_{i2}-\ell \leq n<n_{i2} $,
\begin{eqnarray}
\dfrac{dF_{n}(m=n_{i2},t)}{dt}=W_s(1-P_{ste1})q_{II,III}(\underline{n}|n+\ell)F_{n+1}(m,t)\nonumber \\ -W_s(1-P_{ste1})q_{II,III}(\underline{n}|n+\ell)F_{n}(m,t).\nonumber \\
\label{m_eq_4}
\end{eqnarray}
Not all of the equations from (\ref{m_eq_0})$ - $ (\ref{m_eq_4}) are independent with each other.
For mean first passage time calculation we begin with the initial condition that,
\begin{equation}
F_{m}(m,t)=\delta (t).
\label{boundary}
\end{equation}
To get the analytical expression of $\langle t_{1} \rangle$, and $\langle t_{2} \rangle$, we introduce the Laplace transformations $ \int_{0}^{\infty} e^{-st}F_{n}(m,t)dt $  and with Laplace transformation Eqs. (\ref{m_eq_0})$ - $ (\ref{m_eq_4}) transforms as follows,\\
At site $ n=0 $,
\begin{eqnarray}
sF_{0}((m=n_{i1},n_{i2}),s)=\alpha_{s}P(\underbrace{0,0, \dots, 0}_{\ell})F_{1}(m,s) \nonumber \\ -\alpha_{s}P(\underbrace{0,0, \dots, 0}_{\ell})F_{0}(m,s). \nonumber \\
\label{m_eq_0_lt}
\end{eqnarray}
For $ 1 \leq n<n_{i1}-\ell $,
\begin{eqnarray}
sF_{n}((m=n_{i1},n_{i2}),s)=W_sq_{I,I}(\underline{n}|n+\ell)F_{n+1}(m,s)\nonumber \\ -W_sq_{I,I}(\underline{n}|n+\ell)F_{n}(m,s). \nonumber \\
\label{m_eq_11_lt}
\end{eqnarray}
For $ n_{i1}-\ell \leq n<n_{i1} $,
\begin{eqnarray}
sF_{n}((m=n_{i1},n_{i2}),s)=W_sq_{I,II}(\underline{n}|n+\ell)F_{n+1}(m,s)\nonumber \\ -W_sq_{I,II}(\underline{n}|n+\ell)F_{n}(m,s). \nonumber \\
\label{m_eq_12_lt}
\end{eqnarray}
At site $ n=n_{i1} $,
\begin{eqnarray}
sF_{n}(m=n_{i2},s)=W_sq_{I,II}(\underline{n}|n+\ell)F_{n+1}(m,s)\nonumber \\ -W_sq_{I,II}(\underline{n}|n+\ell)F_{n}(m,s).\nonumber \\
\label{m_eq_2_lt}
\end{eqnarray}
For $  n_{i1}+1 \leq n<n_{i2}-\ell $,
\begin{eqnarray}
sF_{n}(m=n_{i2},s)=W_s(1-P_{ste1})q_{II,II}(\underline{n}|n+\ell)F_{n+1}(m,s)\nonumber \\ -W_s(1-P_{ste1})q_{II,II}(\underline{n}|n+\ell)F_{n}(m,s),\nonumber \\
\label{m_eq_3_lt}
\end{eqnarray}
For $  n_{i2}-\ell \leq n<n_{i2} $,
\begin{eqnarray}
sF_{n}(m=n_{i2},s)=W_s(1-P_{ste1})q_{II,III}(\underline{n}|n+\ell)F_{n+1}(m,s)\nonumber \\ -W_s(1-P_{ste1})q_{II,III}(\underline{n}|n+\ell)F_{n}(m,s),\nonumber \\
\label{m_eq_4_lt}
\end{eqnarray}
and, $ F_{m}(m,s)=1 $.

From Eqs. (\ref{m_eq_0_lt}) $ - $ (\ref{m_eq_4_lt}), we get,
\begin{eqnarray}
F_{0}(n_{i1},s)=\frac{\left(1-\rho_I\right){}^\ell \alpha _{\text{s}} \left[\frac{q_{I,I} W_{\text{s}}}{q_{I,I} W_{\text{s}}+s}\right]^{n_{i1}}}{\left(1-\rho_I\right){}^\ell \alpha _{\text{s}}+s}, 
\label{time_1}
\end{eqnarray}
where, $ \rho_I=\rho_{II} $ and $ q_{I,I}=q_{I,II} $ i.e. both segments \RN{1} and \RN{2} are in same phase.
\begin{widetext}
\begin{eqnarray}
F_{0}(n_{i2},s)=\frac{\left(1-\rho_I\right){}^\ell \alpha _{\text{s}} \left[\frac{q_{I,I} W_{\text{s}}}{q_{I,I} W_{\text{s}}+s}\right]^{n_{{i1}}} \left[\frac{q_{I,I} \left(P_{\text{ste1}}-1\right) W_{\text{s}}}{q_{I,I} \left(P_{\text{ste1}}-1\right) W_{\text{s}}-s}\right]^{n_{\text{i2}}-n_{\text{i1}}}}{\left(1-\rho_I\right){}^\ell \alpha _{\text{s}}+s}, 
\label{time_2}
\end{eqnarray}
where, $ \rho_I=\rho_{II}=\rho_{III} $ and $ q_{I,I}=q_{I,II}=q_{II,II}=q_{II,III} $, i.e., all three segments \RN{1},\RN{2} and \RN{3} are in same phase.

The derivative of Eqs. (\ref{time_1}) and (\ref{time_2}), gives us the mean first passage times,
\begin{eqnarray}
\langle t_1 \rangle = -\dfrac{F_{0}(n_{i1},s)}{ds}\bigg |_{s=0}=\frac{n_{\text{i1}}}{q_{I,I} W_{\text{s}}}+\frac{\left(1-\rho_I\right){}^{-\ell}}{\alpha _{\text{s}}},
\label{time_11}
\end{eqnarray}
and
\begin{eqnarray}
\langle t_2 \rangle =-\dfrac{F_{0}(n_{i2},s)}{ds}\bigg |_{s=0}=\frac{\left(1-\rho_I\right){}^{-\ell} \left[q_{I,I} \left(P_{\text{ste1}}-1\right) W_{\text{s}}+n_{\text{i1}} \left(1-\rho_I\right){}^\ell P_{\text{ste1}} \alpha _{\text{s}}-n_{\text{i2}} \left(1-\rho_I\right){}^\ell \alpha _{\text{s}}\right]}{q_{I,I} \left(P_{\text{ste1}}-1\right) \alpha _{\text{s}} W_{\text{s}}}.
\label{time_22}
\end{eqnarray}
\end{widetext}

\begin{widetext}
\begin{center}
\begin{table}
\begin{center}
\begin{tabular}{| c ||c |c |c |  }
  \hline
  \\
  &\textbf{~~~Low Density (LD)} ~~~& ~~~\textbf{High Density (HD)}~~~ &~~~ \textbf{Maximal Current (MC)}~~~ \\
 \\
 \hline
 \\
  Phase boundary condition &$ \alpha<\beta, ~\alpha<\dfrac{1}{1+\sqrt{\ell}} $& $ \beta<\alpha, ~\beta<\dfrac{1}{1+\sqrt{\ell}} $&$ \alpha >\dfrac{1}{1+\sqrt{\ell}},~\beta>\dfrac{1}{1+\sqrt{\ell}} $\\
 \\
 Flux ($ J $)& $ \dfrac{\alpha(1-\alpha)}{1+\alpha(\ell-1)} $&$ \dfrac{\beta(1-\beta)}{1+\beta(\ell-1)} $ &$\dfrac{1}{(\sqrt{\ell}+1)^2}$\\
  \\
 Coverage density ($ \rho_c $) &$ \dfrac{\alpha\ell}{1+\alpha(\ell-1)} $& $1-\beta$& $\dfrac{\sqrt{\ell}}{(\sqrt{\ell}+1)}$\\
 \\
  \hline
  \end{tabular}
  \caption{Comparison of phase boundary conditions, fluxes and coverage densities in three phases, namely the low density (LD), high density (HD) and maximal current (MC) phases of TASEP of rods corresponding to the forward hopping  probability is $1$.}
  \label{comparison}
  \end{center}
  \end{table} 
  \end{center}
  \end{widetext}

Under steady state condition, each of the five segments I-V can, in principle, exists in one of the three possible phases namely, low density (LD), high density (HD), and maximal current (MC) phase. (See \cite{lakatos03,schad10,shaw03,shaw04b} and Table \ref{comparison}, for comparison of phase boundary conditions, fluxes and coverage densities in three phases). The effective initiation and termination rate constants and their defining expressions for all five segments are given in Table \ref{rate_1}.
\begin{table}
\begin{center}
\begin{tabular}{| c | c | }
  \hline
  Rate constant & Expression  \\
  \hline
  \hline
  $\beta_{eff1} $& $ \beta_{eff1}=q_{I,II}(\underline{n_{i1}}|n_{i1}+\ell) $ \\
  \hline
  $\alpha_{eff2} $& $\alpha_{eff2}=\rho_{0,I}$ \\
  \hline
  $\beta_{eff2} $& $\beta_{eff2}=q_{II,III}(\underline{n_{i2}}|n_{i2}+\ell) $ \\
  \hline
  $\alpha_{eff3} $& $\alpha_{eff3}=\rho_{0,II}+\rho_{1,II}$ \\
  \hline
  $\beta_{eff3} $ & $ \beta_{eff3}= \beta_{1}+(1-\beta_{1})q_{III,IV}(\underline{n_{s1}}|n_{s1}+\ell)$ \\
  \hline
  $\alpha_{eff4} $& $\alpha_{eff4}=\rho_{0,III}+(1-\beta_1)\rho_{1,III}+\rho_{2,III}$ \\
  \hline
  $\beta_{eff4} $ & $ \beta_{eff4}= \beta_{2}+(1-\beta_{2})q_{IV,V}(\underline{n_{s2}}|n_{s2}+\ell)$ \\
  \hline
  $\alpha_{eff5} $& $\alpha_{eff5}=\rho_{0,IV} +(1-\beta_2)\rho_{2,IV}$ \\
  \hline
  \end{tabular}
  \caption{The effective rate constants and their defining expressions.}
  \label{rate_1}
  \end{center}
  \end{table} 

\begin{figure}[t]
\begin{center}
\includegraphics[angle=0,width=0.9\columnwidth]{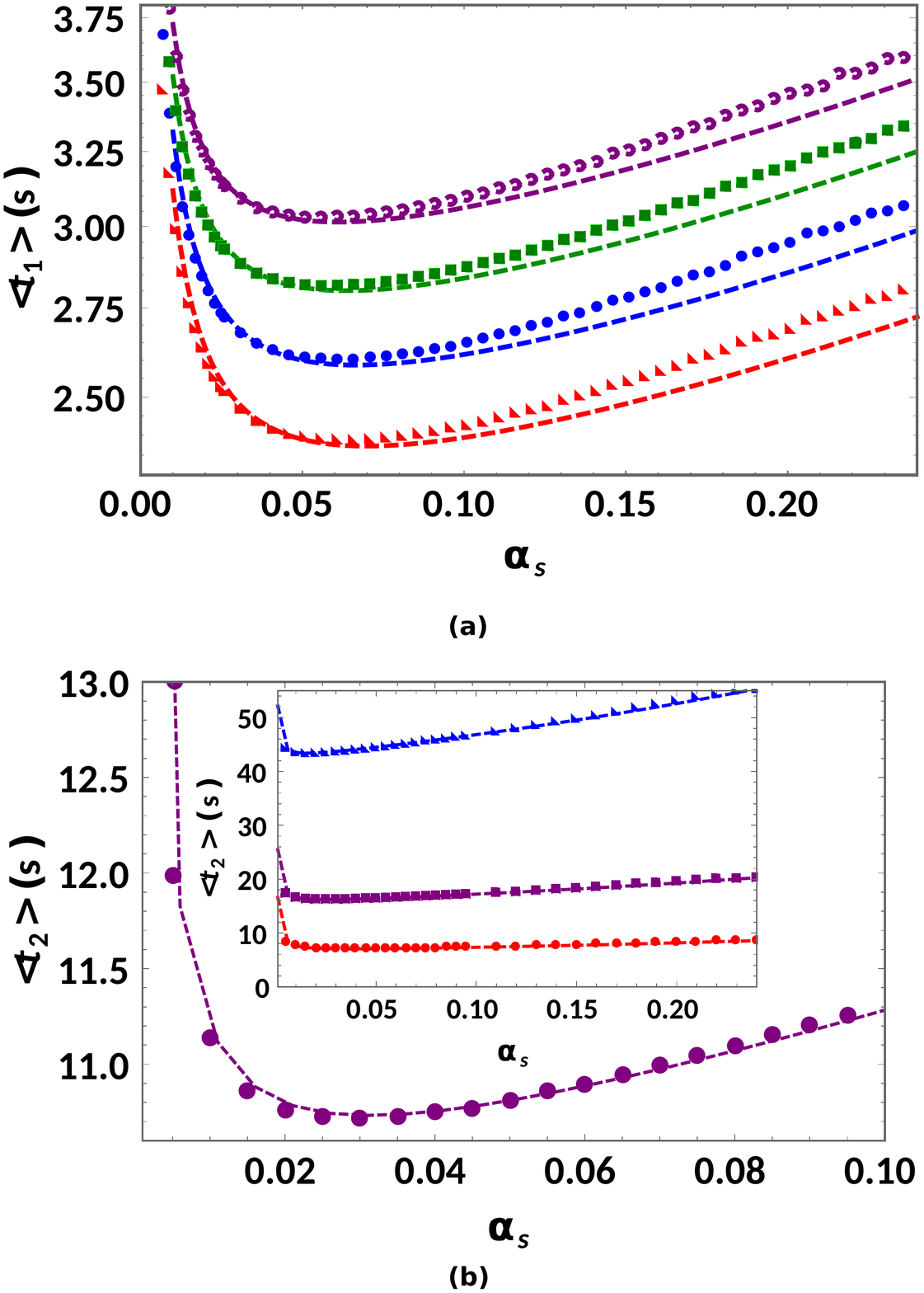}
\end{center}
\caption{(Color online) MFPTs $\langle t_1 \rangle~\text{and}~\langle t_2 \rangle $ are plotted against the rate of attachment  $ \alpha_s $, for different values of $ P_{ste1} $. In inset of (b) we show the variation in $ \alpha_s $ from $ 0 $ to the maximum critical value of $ \alpha_s $ (below which system remains in $LD|LD|LD|LD|LD$ phase i.e. $ \alpha_s=0.24 $, while in main figure we show only a small range of $ \alpha_s $ to see the dominant effect of crowding. The theoretical predictions obtained under MFA are shown by dashed curves, while the numerical data obtained from Monte-Carlo simulations are shown by dicrete symbols. In (a) triangle, disk, square and circle correspond to $ UTR_1=200 $, $ UTR_1=220 $, $ UTR_1=240 $, and $ UTR_1=260 $, respectively. In (b) in main figure result is shown for $ P_{ste1}=0.1 $ and $ UTR_1=200 $, whereas in inset, disk, square and triangle correspond to $ P_{ste1}=0.1 $, $ P_{ste1}=0.7 $, $ P_{ste1}=0.9 $, respectively and $ UTR_1=200 $.  The numerical values of other parameters used here are, $ \beta_s=0.8, P_{ste2}=0.7,W_s=1.0,W_{e1}=1.0,W_{e2}=1.0,\beta_1=0.8,\beta_2=0.8,\ell=10 $, $\text{UTR}_2=n_{i2}=600$.}
\label{fig-res_1}
\end{figure}
\begin{figure}[t]
\begin{center}
\includegraphics[angle=0,width=0.9\columnwidth]{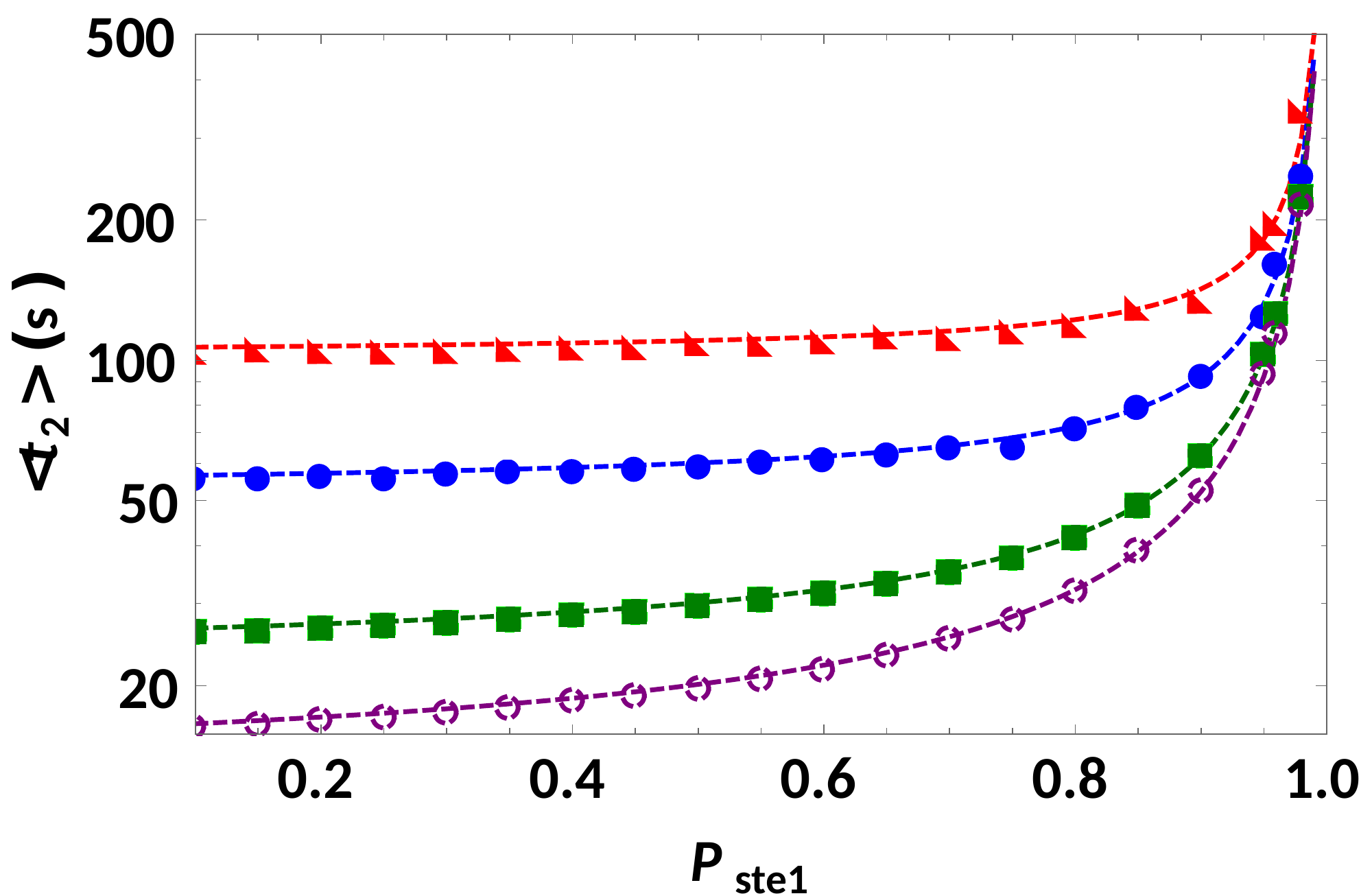}
\end{center}
\caption{(Color online) MFPTs $\langle t_2 \rangle $ are plotted with $P_{ste1}$ for three different values of scanning ribosome attachment rate $ \alpha_{s} $. In inset results are plotted for a small range of $P_{ste1}$ to see the dominant effect of crowding.
The theoretical predictions obtained under MFA are shown by dashed curves, while the numerical data obtained from Monte-Carlo simulations are shown by discrete symbols. Circle, square, disk and triangle correspond to $ \alpha_s=0.001 $, $ \alpha_s=0.0005 $, $ \alpha_s=0.0002 $, and $ \alpha_s=0.0001 $, respectively. The numerical values of other parameters used here are, $ \beta_s=0.8, P_{ste2}=0.7,W_s=1.0,W_{e1}=1.0,W_{e2}=1.0,\beta_1=0.8,\beta_2=0.8,\ell=10 $, $ \text{UTR}_1=n_{i1}=200~\text{and}~\text{UTR}_2=n_{i2}=600$. }
\label{fig-res_2}
\end{figure}
%

\begin{figure}[t]
\begin{center}
\includegraphics[angle=0,width=0.8\columnwidth]{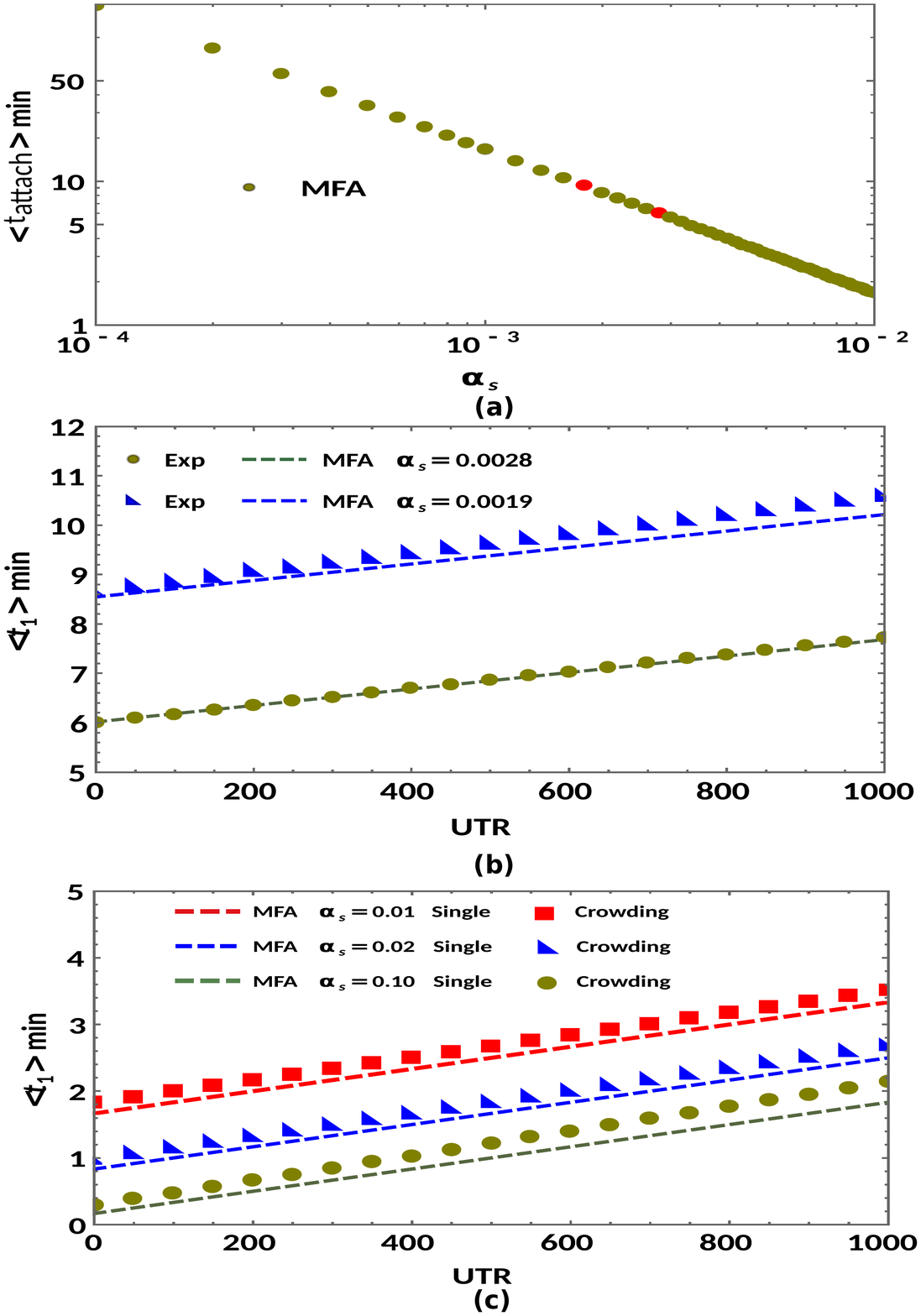}
\end{center}
\caption{(Color online) (a) Variation in $ \langle t_{attach} \rangle $ 
with the rate of attachment $ \alpha_s $. Theoretical results marked by two red dots are quantitatively comparable with the experimental data, extracted from ref.\cite{vassilenko11a}. The results in (b) have been obtained in the special case where at any instant only one scanning rod0 is allowed to search for the start site.  The MFPTs $\langle t_1 \rangle$ are plotted against the length of UTR, for two different values of $ \alpha_s $. The theoretical predictions obtained under MFA are shown by dashed curves, while the experimental data extracted from \cite{vassilenko11a} are shown by dicrete symbols. The numerical values of the other parameters used here are, $ \beta_s=0.8, P_{ste2}=0.7,W_s=0.1,W_{e1}=0.1,W_{e2}=0.1,\beta_1=0.8,\beta_2=0.8~\text{and}~\ell=10 $. In (c) MFPTs $\langle t_1 \rangle$ are plotted against the length of UTR, for three different values of $ \alpha_s $. 
Results for the case of a single rod0 are shown by dashed curves, while the predictions under the crowding rods are shown by discrete symbols.
All the results are obtained analytically under MFA.}
\label{fig-res_3}
\end{figure}


\begin{figure}[t]
\begin{center}
\includegraphics[angle=0,width=0.8\columnwidth]{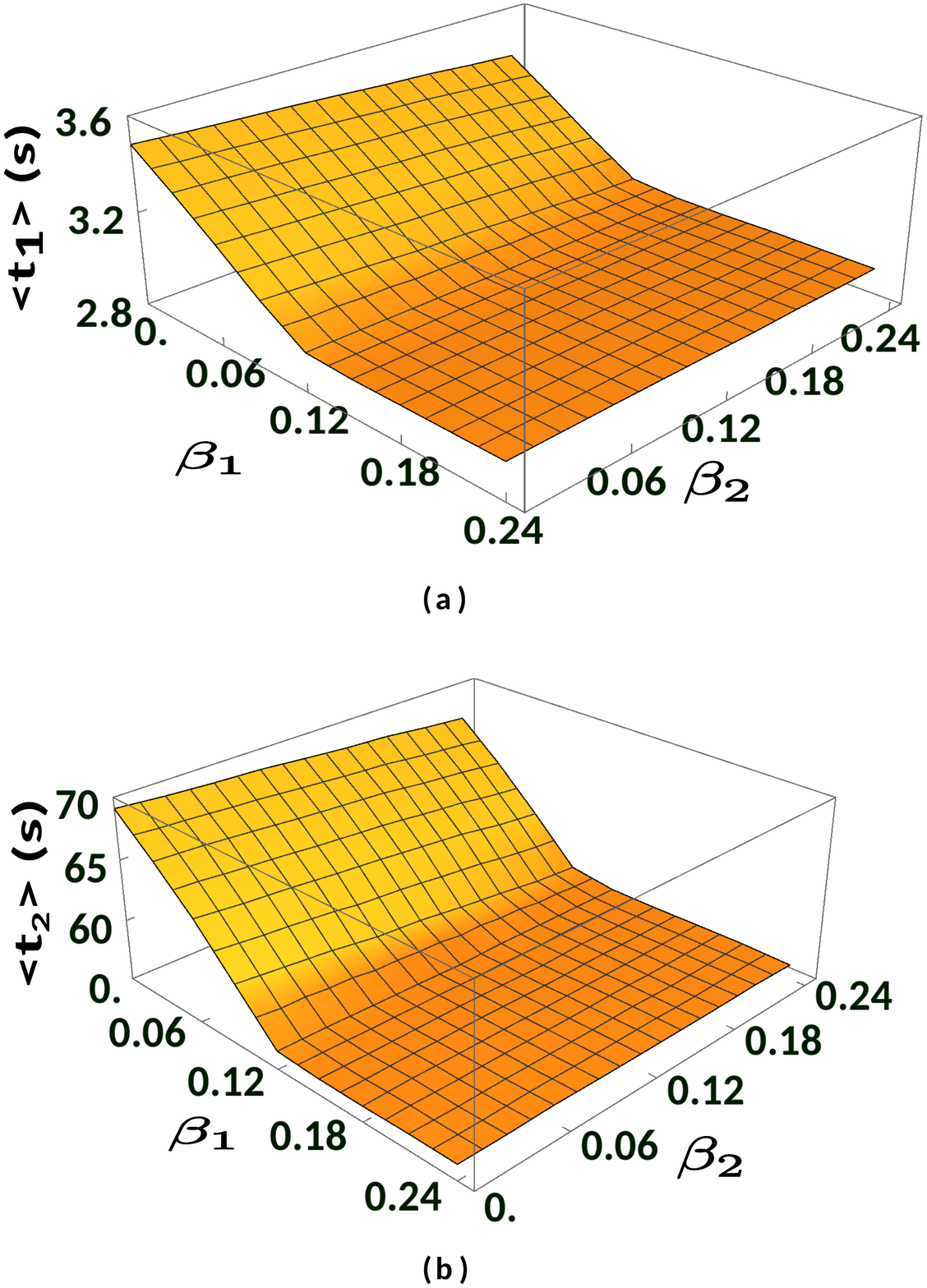}
\end{center}
\caption{(Color online) In (a) 3D plot of MFPT $\langle t_1 \rangle$ and in (b) 3D plot of MFPT $\langle t_2 \rangle$ are plotted against the rate of detachments  $ \beta_1~\text{and}~\beta_2 $ (for constant values of 
$\alpha_{s}=0.8,P_{ste1}=P_{ste2}=0.9~\text{and}~\beta_{s}=0.15$). Over the entire range of both $ \beta_1~\text{and}~\beta_2 $ (from $ 0 $ to $ 0.24 $) in this Figure, the system remains in $HD|HD|HD|HD|HD$ phase.
Results shown here are obtained from the MC simulations. The numerical values of other parameters used here are, $W_s=1.0,W_{e1}=1.0,W_{e2}=1.0,\ell=10 $, $\text{UTR}_2=n_{i2}=600$.}
\label{fig-res_4}
\end{figure}
\color{black}
\begin{figure}[t]
\begin{center}
\includegraphics[angle=0,width=0.9\columnwidth]{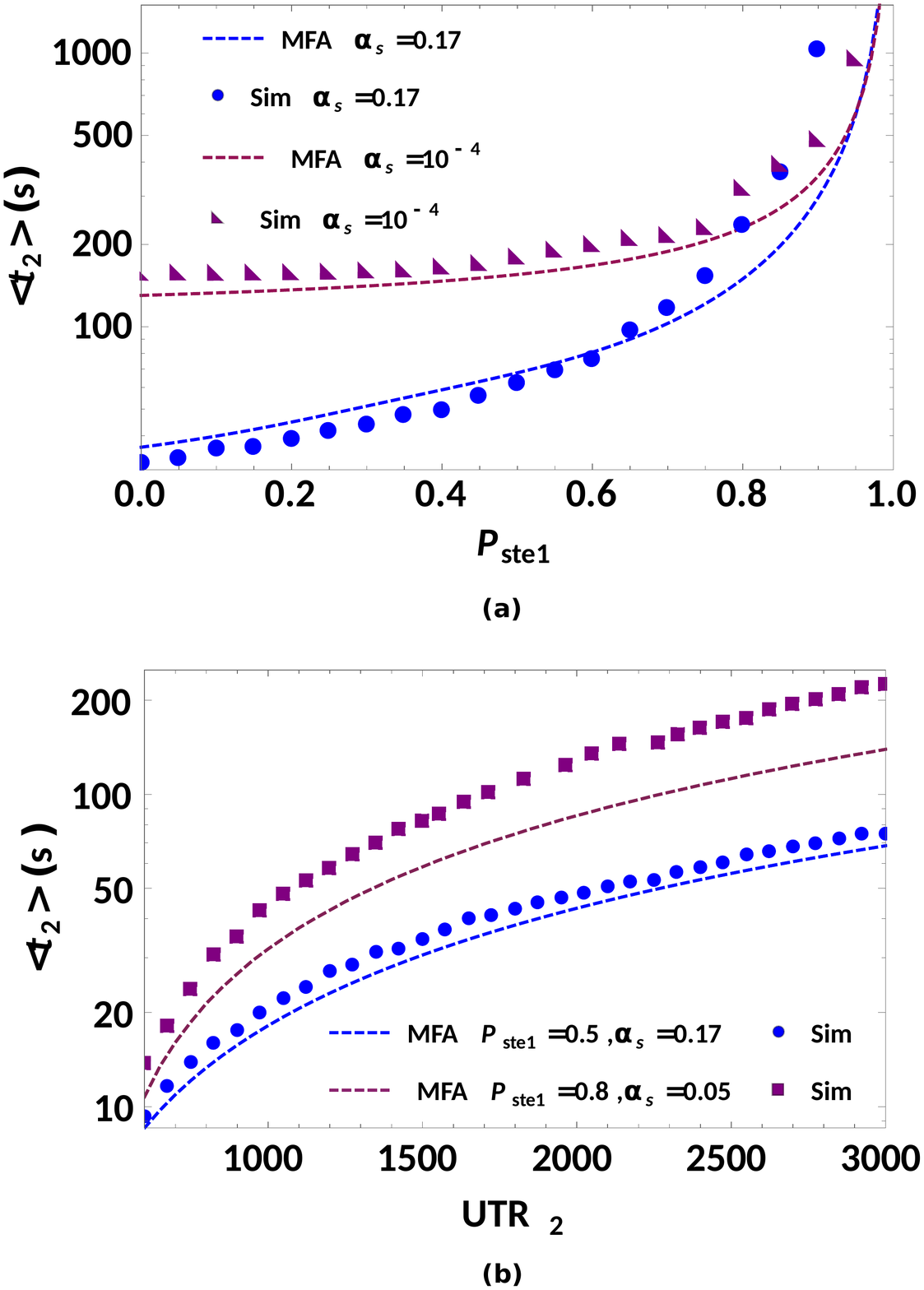}
\end{center}
\caption{(Color online) In (a) MFPTs $\langle t'_2 \rangle $ are plotted against $ P_{ste1} $, for two different values of $ \alpha_s $; here $ k_{off}=0.5 $, $ n_{i2}-n_{i1}=2500 $. In (b) we display semi Log plots of MFPTs $\langle t'_2 \rangle $  against the length of segment \RN{2} (which has been defined as UTR$_{2}$). The theoretical predictions obtained under MFA are shown by dashed curves, while the numerical data obtained from Monte-Carlo simulations are shown by dicrete symbols. The numerical values of other parameters used here are, $ K_{off}=0.5,\beta_s=0.8, P_{ste2}=0.7,W_s=1.0,W_{e1}=1.0,W_{e2}=1.0,\beta_1=0.8,\beta_2=0.8,\ell=10 $, $ \text{UTR}_1=n_{i1}=500~\text{and}~\text{UTR}_2=n_{i2}=3000$.}
\label{fig-res_5}
\end{figure}
There are three key parameters of the system affecting the MFPTs, (a) effective  rate of entry of a new rod0 at the initiation site $ i=1 $, i.e. $ \alpha_sP(\underbrace{0,0,\dots,p}_{\ell}) $, (b) effective elongation rate of rod0, i.e. $ W_sq_{\mu,\nu}(i|\underline{i+\ell})$, and (c) length of UTR .

Since the system consists of five segments and each segment can exist in three possible phases. Ideally, the entire system can be found in $ 3^5 $ composite phases. However, not all of the $ 3^5 $ phases are physically unattainable because of the symmetry requirements and steady-state conditions. Some of them are biologically irrelevant even when they are physically attainable. In our notation the composite phase of the system is denoted by  $X|Y|Z|Z|X$, here $ X, Y~\text{and}~Z $ is either LD, HD and MC phase. In this work, we show the results only for few composite phases. However, our results can be extended for any physically attainable composite phase by using the corresponding expressions of number density in various phases.
\subsubsection{Variation of MFPTs with $ \alpha_s $}
In Fig. \ref{fig-res_1}, we show the variation in MFPTs $ \langle t_1 \rangle $ and $ \langle t_2 \rangle $ with the initiation rate 
$\alpha_s$. Here, we show the result for the composite phase $ LD|LD|LD|LD|LD $. MFPTs exhibit a nonmonotonic behavior as we vary $ \alpha_s $. In order to explain the physical origin of this nonmonotonicity, we draw attention to the fact that the total time needed to reach a given target site on the lattice is the sum of two contributions: (a) the time required for a rod to attach with the site $i=1$ (i.e., to make the entry into the system through $i=1$), and (b) the time needed to travel from the site of entry to the target site. At sufficiently small values of $ \alpha_s $, entry of a rod is the rate-limiting process, i.e., the time needed to enter the lattice is much longer than the time needed to walk from the entry to the target site. Therefore, in this regime, the larger is the value of $ \alpha_s $, the shorter is the time needed to enter the lattice, thereby resulting in a lowering of the corresponding MFPT. However, increasing $\alpha_{s}$ also increases the number density of the rods which causes mutual hindrance against their forward movement that, in turn, results in a long time of travel from the entry to the target site. Therefore, beyond a certain value of $ \alpha_s $, the adverse effect of crowding becomes so strong that the resulting increase in the time of travel along the lattice cannot be compensated by the reduction in the time required for the entry of the rod; therefore, at sufficiently large values of $\alpha_{s}$, its further increase causes only overall increase  of the  MFPTs. 

The nonmonotonic variation of $ \langle t_1 \rangle $ and $ \langle t_2 \rangle $ should not be confused with the non-monotonic variation of the flux of rods in a single-species TASEP with increasing rod density. Crowding, i.e., traffic congestion, is the cause of both the increase of the MFTs with increasing $\alpha_{s}$ beyond the minimum MFT and the decrease of flux with increasing particle density beyond the maximum value of the flux. However, the physical origin of decrease of the MFTS at low-$\alpha_{s}$ and that of the increase of the flux at low rod density are completely different. In the low-$\alpha_{s}$ regime, increase in 
$\alpha_{s}$ shortens the waiting time before the actual entry of a rod thereby reducing the corresponding MFT. In contrast, in the low-density regime of a single-species TASEP, the increase of flux with increasing number density trivially follows from the definition that flux is proportional to the number density.

\subsubsection{Variation of MFPTs with $P_{ste1}$}

The variation of MFPT $\langle t_2 \rangle$ with $ P_{ste1} $  is  shown in Fig. \ref{fig-res_2} for a few different values of $ \alpha_s $.  As long as $P_{ste1}$ is sufficiently small, practically all the rod0 miss the site $i=n_{i1}$ and continue searching beyond this site. Increasing $P_{ste1}$ depletes the population of scanning rod0 in the segment II leading to a longer wait till a rod0 successfully hits the target. The trend of variation with $\alpha_{s}$ is also consistent with intuitive expectation: the smaller is the value of $\alpha_{s}$ the longer is the MFPT $\langle t_{2} \rangle$. As $P_{ste1}$ approaches its maximum allowed value $1$, the value of $\langle t_{2} \rangle$ rises sharply because very rarely a rod0 escapes conversion to rod1 and, eventually, hits the target at $i=n_{i2}$. 

\subsubsection{Variation in MFPTs with the length of UTR} 

Only a few quantitative experimental results have been reported in the literature with which comparison of our theoretical predictions is possible. Fortunately, the variation of the MFPTs with the lengths of UTRs provide an opportunity to make a direct connection with experimental data. One set of such experimental data are extracted from \cite{vassilenko11a} in Fig. \ref{fig-res_3} (a), we show the results for a special case where at any instant only a single scanning rod0 searches for the start site in the absence of crowding. The two red dots in this figure are special points where the attachment time of a new rod0, i.e., $ \langle t_{attach} \rangle$ is quantitatively comparable with the experimental data, given in \cite{vassilenko11a}. Therefore, to compare our theoretical findings with the corresponding experimental results,  we use only these two values of the attachment rate, i.e., $ \alpha_s=0.0028~\text{and}~0.0019 $ in Fig. \ref{fig-res_3} (b),  we show the variation in MFPT for a special case where at any instant only one scanning rod0 searches for the start codon in the absence of crowding. In Fig. \ref{fig-res_3} (c)  we show the effect of crowding on MFPT as we vary the length of UTR.  

\subsubsection{Variation in MFPTs with the detachment rates $ \beta_1 $ and $\beta_2$} 
For small values of detachment rates $ \beta_1 $, $\beta_2$ and $\beta_{s}$ along with $ \alpha_{s} \geq 0.24 $, for $ \ell=10 $, system achieves the high density (HD) phase in all five segments. Here, detachment of rod1 and rod2 from their respective stop codon sites $ n_{s1} $ and $ n_{s2} $ with rates $ \beta_1 $ and $ \beta_2 $, respectively and rod0 from the last site of the lattice with rate $ \beta_{s} $, are the rate limiting steps and regulates the MFPTs $\langle t_1 \rangle~\text{and}~\langle t_2 \rangle$. In Fig. \ref{fig-res_4}, we show the 3D plots of $\langle t_1 \rangle~\text{and}~\langle t_2 \rangle$ with $ \beta_1 $ and $\beta_2$. In this figure, below a critical value of $ \beta_1 $, MFPTs decrease with the increase in $ \beta_1 $. If we further increase $ \beta_1 $ MFPTs saturate and achieve a constant value. However, MFPTs decease with $ \beta_2 $ only for the small values of $ \beta_1 $.  These  remain practically unchanged with further increase of $ \beta_2 $ for large values of $ \beta_1 $.  

\subsubsection{Variation of MFPTs with $k_{off}$}

In our basic model scanning rods are not allowed to detach from the lattice while searching for the start site. However, it is also experimentally verified fact that a scanning rod0 (SSU) might detach from the lattice once it encounters another species of rod (ribosome engaged in protein synthesis).  In order to capture this additional feature, we propose that, if there is a rod0 at site $ i $, it can detach from the lattice with rate $ K_{off} $, provided the target location $ i+\ell $ is occupied by another rod of different species (See Fig. \ref{fig-model_1}(b)). The occupational probability of scanning rod0 is maximum in segment \RN{1} and then in segment \RN{2}, and further, it falls as we proceed further to the segments \RN{3} and \RN{4}. Consequently, the probability of an encounter of a scanning rod0 with another rod of different species is maximum inside segment \RN{2}, and it is negligible inside segments \RN{3} and \RN{4}. Therefore, in this model, we ignore the detachment of a rod0 from segments \RN{3,} and \RN{4} and detachment is allowed only inside segment \RN{2}. For this extended version of the model, we measure only $ \langle t_2 \rangle $, because $ \langle t_1 \rangle $, remains unaffected by the detachment of scanning rod0.


In the stationary state, for a given set of $ k_{off}, P_{ste1}~ \text{and} ~\alpha_s $, the number of rod1s, inside segment \RN{2} is given by,
\begin{eqnarray}
N_{e1}=\rho_{1,II}(n_{i2}-n_{i1}), \\ \nonumber
\rho_{1,II}=P_{ste1}\rho_{I}, \\ \nonumber
\rho_{I}=\dfrac{\alpha_{s}}{1+\alpha_{s}(\ell-1)},
\label{low_dens_number_e1}
\end{eqnarray}
if both segments \RN{1} and \RN{2} are in low density (LD) phase.  Similarly, we can write the expression of $ N_{e1} $ for all other phases. 

In general, a rod0 inside segment \RN{2} encounters $ N_{e1} $ rod1s before it reaches the second start codon site. In an encounter with rod1, they might detach from the lattice with an effective detachment probability, $ K'_{off} $, which is given by 
\begin{equation}
K'_{off}=K_{off}p, 
\end{equation}
 or survives with probability $1- K'_{off} $.
Here, $ p $ is defined as the probability of finding a rod1 at site $ i $ provided there is a ribosome (either, a rod0 or a rod1) at site $ i-\ell $ and it is given by,
\begin{equation}
p=\dfrac{\rho_{1,II}}{1+\rho_{1,II}-\rho_{1,II}\ell}.
\label{m2_rhoe2}
\end{equation} 

To reach the second start site a rod0 must survive all $ N_{e1} $ encounters with rod1s. Therefore, their effective survival probability after $ N_{e1} $ encounters is given by,
\begin{equation}
P_{s}=(1-K'_{off})^{N_{e1}}=(1-K'_{off})^{\rho_{1,II}(n_{i2}-n_{i1})}.
\label{eff_sur_1}
\end{equation}

We also know that the number of rod0s inside segment \RN{2} in the absence of detachment i.e $ K_{off}=0 $ will be,
\begin{eqnarray}
N_{s}=\rho_{0,II}(n_{i2}-n_{i1}), \nonumber \\ 
\rho_{0,II}=(1-P_{ste1})\rho_{I},  \nonumber \\
\rho_{I}=\dfrac{\alpha_{s}}{1+\alpha_{s}(\ell-1)},
\label{low_dens_number_e2}
\end{eqnarray}
if both segments \RN{1} and \RN{2} are in low density (LD) phase. (Similarly, we can write the expression of $ N_{s} $ for all other phases.) By combining Eqs. (\ref{eff_sur_1}) and (\ref{low_dens_number_e2}), we find that the number of rod0s which survive after $ N_{e1} $ encounters is given by
\begin{equation}
 N_{s_{sur}} \approx \text{Int}[N_{s}P_{s}] \approx N_{s}\big[(1-K'_{off})^{\rho_{1,II}(n_{i2}-n_{i1})}\big].
 \label{survived_number_1}
 \end{equation} 
Similarly, number of detached rod0s before the arrival of first rod0 at the second start site is given by 
\begin{equation}
 N_{s_{detach}} \approx N_{s}-\text{Int}[N_{s}P_{s}] \approx N_{s}\Big[1-\big[(1-K'_{off})^{\rho_{1,II}(n_{i2}-n_{i1})}\big]\Big].
 \label{detached_number_1}
 \end{equation} 
By combining the Eqs. (\ref{time_22}) and (\ref{detached_number_1}), we get the MFPT $ \langle t'_2 \rangle $ to be
\begin{equation}
\langle t'_2 \rangle \approx  N_{s_{detach}}(1/\alpha'_{s}) + \langle t_2 \rangle
\label{time_detach_mfpt}
\end{equation}
where, $ \alpha'_{s}=\alpha_{s}(1-\rho_{0,I})^{\ell} $ and $ \langle t_2 \rangle $ is the MFPT for the second start site in the absence of detachment, i.e. $ K_{off}=0 $. 
In Fig. \ref{fig-res_5}(a), we show the variation in MFPT $\langle t'_2 \rangle$ with $ P_{ste1} $ and in (b) with the length of UTR for various values of detachment rate $ K_{off} $. 

\section{Summary and Conclusion}

In this paper, we have developed a biologically motivated one-dimensional model of a three-species exclusion process. The lattice, which represents an mRNA, consists of five segments labeled by the index I-V. One species of rods, designated rod0, is unique in the sense that it can get probabilistically converted into rod1 and rod2 species at two special sites $n_{i1}$ and $n_{i2}$ marked at the beginning of the segments II and III, respectively. The ends of the segments III and IV labeled as $n_{s1}$ and $n_{s2}$,  are also special in that at these sites the rod1 and rod2, respectively, either make an exit from the lattice or get re-converted back to rod0. Consequently, interference of the flow of different species of rods takes place over the three segments II, III and IV. Each rod0 represents the SSU of a ribosome whereas rod1 and rod2 represent fully assembled ribosomes engaged in the synthesis of two distinct proteins encoded by overlapping genes 1 and 2, respectively.   

The {\it first time} any of the members of the common pool of free rods succeeds in reaching the site $n_{i1}$ ($n_{i2}$) and gets converted to a rod1 (rod2) is identified as the time taken to initiate the synthesis of a protein of type 1 (type 2). Thus, be definition, these times of initiation of protein synthesis are {\it first-passage} times. We have analytically calculated the mean of these first-passage times using the formalism of backward master equation. The results of this approximate analytical theory are in reasonably good agreement with the corresponding numerical data that we have obtained by carrying out Monte-Carlo simulations of the model. We have also made the quantitative connection of our theory with whatever little experimental data could be extracted from the published literature \cite{vassilenko11a}. We have reported several new theoretical predictions that, we hope, can be tested experimentally in the near future.

\section*{Acknowledgements} 

This work has been supported by a J.C. Bose National Fellowship (DC) and by UGC Senior Research Fellowship (BM). DC also thanks to the Visitors Program of the Max-Planck Institute for the Physics of Complex Systems for hospitality during the final stages of the preparation of this manuscript. 


\end{document}